\documentclass[a4paper]{article}
\usepackage[utf8]{inputenc}
\usepackage[T1]{fontenc}
\usepackage{amsmath,amssymb,amsfonts}
\usepackage{algorithmic}
\usepackage{graphicx}
\usepackage{xcolor}
\usepackage{subcaption}
\usepackage[sort]{cite}
\usepackage{epstopdf}
\usepackage{url}
\usepackage{enumerate}
\usepackage{physics}

\title{Analytic Modelling of a Simplified Planar Goubau Line}
\author{Tobias Schaich, Daniel Molnar, Anas Al Rawi, Mike Payne}
\date{\today}

\begin{document}
	\maketitle
	\begin{abstract}
		This paper analyses the surface wave mode propagating along a simplified planar Goubau line consisting of a perfectly conducting circular wire on top of a dielectric substrate of finite thickness but infinite width. An approximate equation for the propagation constant is derived and solved through numerical integration. The dependence of the propagation constant on various system parameters is calculated and the results agree well with full numerical simulations. In addition, the spatial distribution of the longitudinal electric field is reported and excellent agreement with the numerical simulation and previous studies is found. Moreover, validation against experimental phase velocity measurements is also reported. Finally, insights gained from the model are considered for a Goubau line with a rectangular conductor. These results present the first step towards an analytic model of the planar Goubau line.
	\end{abstract}
	
	\section{Introduction}
	Surface waves (SW) on circular conducting wires have been of theoretical interest since their discovery by Sommerfeld in 1899 \cite{Sommerfeld1899}. However, due to the large lateral extent of the fields at low frequencies, practical applicability seemed limited at first. Goubau discovered that by coating wires in dielectric or corrugating the wire's surface, the fields' lateral confinement could be drastically enhanced \cite{Goubau1950}. These coated wires, named Goubau lines in later years, showed low loss and weak dispersion. Hence, they were discussed as an alternative to traditional, two-conductor transmission lines. Recently, interest in surface wave technology has re-emerged at the GHz-THz frequency range where it presents a promising alternative to current waveguide technology \cite{Dfg2004,Jeon2005}. Furthermore, by introducing sub-wavelength corrugations, spoof surface plasmon polaritons emerge which have tunable properties and can exhibit sub-wavelength lateral confinement \cite{Pendry2004, Shen2013, Tang2018}. These new technologies have been discussed as solutions to problems such as signal integrity in integrated circuits and backhaul solutions for the network standard 5G \cite{Zhang2015,Galli2018}.
	
	In many cases it is favourable to print a conductor design on a substrate using established printed circuit board fabrication processes such as etching. This led to the invention of a planar Goubau line (PGL) consisting of a thin rectangular conducting strip on a substrate \cite{Akalin2006,Treizebre2005, Gacemi2013, Tang2017 }. Multiple electronic components have been proposed for PGLs including broadband loads, power dividers and frequency selective filters \cite{Xu2011, Chen2011, Horestani2013}. Additionally, application of PGLs in terahertz spectroscopy has been established \cite{Russell2013}.
	
	Despite these advances, only numerical and experimental studies have been published on the PGL to date and no analytic theory or model exists \cite{Xu2006, Gacemi2013a}. A difficulty in the exact treatment is presented by the presence of sharp corners which introduce lightning rod effects as reported in Ref. \cite{Gacemi2013}. Therefore, a simplified model of the Goubau line in which the rectangular conductor is exchanged for a conductor with circular cross section will be considered in this paper. For ease of notation the simplified system will be referred to as a PGL as well. A related system - the single conductor above a semi-infinite conducting earth - has been extensively studied (see \cite{Olsen2000} and references therein). We will draw parallels to this system where appropriate. 
	
	The paper is structured as follows: First, we discuss the wave created by an infinitesimally small current filament above a substrate. Then, the finite thickness of wire is incorporated to derive a characteristic equation for the system and applicability criteria for this approach are discussed. Using the derived equation, the dependence of the propagation constant on input parameters is established and some field patterns are reported. We validate our results against numerical data obtained through the finite element method and experiment. Finally, we draw parallels to the Goubau line with rectangular conductor. 
	
	\section{Derivation of the Electromagnetic Fields}
	
	\begin{figure}[t]
		\centering
		\includegraphics[trim={0 40 0 100}, clip,width=0.9\textwidth]{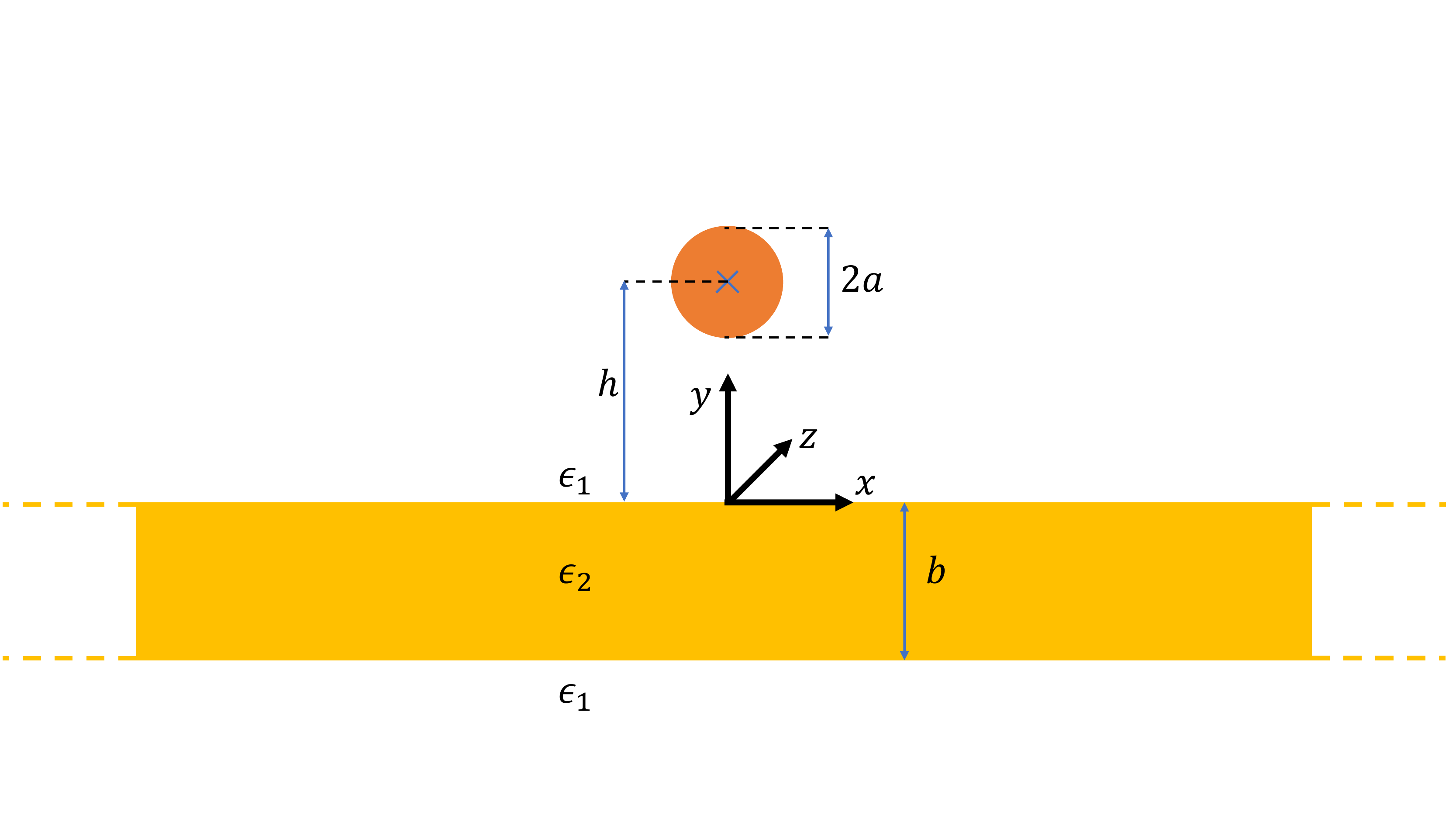}
		\caption{Cross sectional illustration of the simplified planar Goubau line. A wire of radius $a$ is at a height $h$ over a dielectric slab with permittivity $\epsilon_2$ and thickness $b$. The surrounding medium has permittivity $\epsilon_1$. A cartesian coordinate system $(x,y,z)$ is defined with its origin placed on the substrate in line with the centre of the conductor.}
		\label{fig:setup}
	\end{figure}
	
	The system we are investigating consists of a perfectly conducting wire of radius $a$ with its centre located at a height $h$ above a substrate of thickness $b$. Figure \ref{fig:setup} shows the cross section of the system and defines the coordinate system $(x,y,z)$ located on the surface of the substrate with its origin in line with the centre of the wire. We assume the wire and substrate are uniform in the $z$-direction which will be the direction of wave propagation. Additionally, the substrate extends infinitely far in the $x$-plane. It has a dielectric constant of $\epsilon_2$ and is immersed in a medium with dielectric constant $\epsilon_1$ with $\epsilon_1<\epsilon_2$. For most practical applications the surrounding medium is air whose dielectric constant can be approximated as the dielectric permittivity of vacuum $\epsilon_0$. All materials are assumed to be non-magnetic and have permeability equal to the magnetic constant $\mu_0$. 

	We start our derivation by postulating a time harmonic current density $\vec{J}$ that replaces the wire. Its form, which is motivated by the current found in regular surface waveguides such as the Goubau line, is given by:
	\begin{equation}
		\vec{J}=I\delta(x)\delta(y-h)e^{i\omega t - i\beta z}\hat{z}
	\end{equation}
	with angular frequency $\omega$, time $t$, propagation constant $\beta$, current amplitude $I$, unit vector $\hat{z}$ and Dirac delta function $\delta$. This approach is only strictly valid in the case where the wire is infinitesimally small but it can also give reasonable results for thin wires. The physical conditions under which wires may be considered thin will be given later when discussing the characteristic equation. The presented current density can be interpreted as the exact current distribution averaged across the cross section of the wire. For now, we note that as we are not resolving the exact current distribution inside the wire, the near field close to the wire will deviate from an exact solution. However, at distances much greater than the wire radius, the distribution of current inside the wire should have an insignificant effect on the electromagnetic fields. 
	
	Next, we assume that the total field can be separated into a transverse magnetic (TM) and transverse electric (TE) component which are characterised by having no longitudinal magnetic or electric field, respectively. Thus, we may express the total electric and magnetic field,  $\vec{E}$ and $\vec{H}$, as 
	\begin{align}
		\vec{E}=\vec{E}^{TM}+\vec{E}^{TE} &&
		\vec{H}=\vec{H}^{TM}+\vec{H}^{TE}
	\end{align} 
	We expect the fields to be in phase with the current density and should also contain a factor $e^{i\omega t - i \beta z}$ which will be implicitly assumed but omitted for clarity. Faraday's and Ampere's law then take the form
	\begin{align}
		\vec{\nabla} \times \vec{H}^{TM}&=-i\omega \epsilon_j\vec{E}^{TM}+I\delta(x)\delta(y-h)\hat{z}  	 	&\vec{\nabla} \times \vec{E}^{TM}&=i \omega \mu_0 \vec{H}^{TM} \label{eq:Maxwell-TM} \\
		\vec{\nabla} \times \vec{H}^{TE}&=-i\omega \epsilon_j \vec{E}^{TE}&\vec{\nabla} \times \vec{E}^{TE}&=i \omega \mu_0 \vec{H}^{TE} \label{eq:Maxwell-TE}
	\end{align}
	where $\epsilon_j$ takes either the value $\epsilon_1$ or $\epsilon_2$. From the defining property of TE and TM modes, it may be shown that all field components may be calculated from the longitudinal component of the magnetic and electric field, $H_z$ and $E_z$, respectively \cite{Jones1964}. Taking the curl of the second equation in \eqref{eq:Maxwell-TM} and using Faraday's and Gauss' law in combination with the continuity equation as well as standard vector calculus identities, we arrive at 
	\begin{equation}
		\Big(\frac{\partial^2}{\partial x^2}+\frac{\partial^2}{\partial y^2}+\gamma_j^2\Big)E_z^{TM}=-i\omega \mu_0I\frac{\gamma_j^2}{k_j^2}\delta(x) \delta(y-h) \label{eq:Helmholz-TM}
	\end{equation} 
	where we have introduced $k_j^2=\omega^2 \epsilon_j \mu_0$ and the new variable $\gamma_j^2=k_j^2-\beta^2$ with $j\in\{1,2\}$. In a similar fashion, we may manipulate equation \eqref{eq:Maxwell-TE} to arrive at 
	\begin{equation}
		\Big(\frac{\partial^2}{\partial x^2}+\frac{\partial^2}{\partial y^2}+\gamma_j^2\Big)H_z^{TE}=0 \label{eq:Helmholtz-TE}
	\end{equation}
	As the boundaries along the dielectric substrate extend infinitely along the x-direction, it is convenient to introduce a Fourier transform and its inverse as
	\begin{align}
		E^{TM}_z=\frac{1}{2\pi}\int_{-\infty}^{\infty}\tilde{E}^{TM}_z e^{-i \xi x}d\xi	&&\tilde{E}^{TM}_z=\int_{-\infty}^{\infty}E_z^{TM} e^{i \xi x} dx \label{eq:Fourier}
	\end{align}
	Similarly, $\tilde{H}_z^{TE}$ is the Fourier transform of $H_z^{TE}$. Generally, we signify functions in Fourier space by a tilde. The transformed equations \eqref{eq:Helmholz-TM} and \eqref{eq:Helmholtz-TE} become:
	\begin{gather}
		\Big(\frac{\partial^2}{\partial y^2}+u_j^2\Big)\tilde{E}_z^{TM}=-i\omega \mu_0I\frac{\gamma_j^2}{k_j^2} \delta(y-h)\\
		\Big(\frac{\partial^2}{\partial y^2}+u_j^2\Big)\tilde{H}_z^{TE}=0
	\end{gather}
	with $u_j^2=\gamma_j^2-\xi^2$. Solutions to these equations are readily available \cite{Barton1989}. We impose the condition that fields should decay towards infinity and find 
	\begin{align}
		\tilde{E}^{TM}_z=&
\begin{cases}
-\omega \mu_0 I \frac{\gamma_1^2}{k_1^2} \frac{e^{i u_1 |y-h|}}{2u_1}+C_1 e^{iu_1y}  \qquad y\geq0\\
C_2 e^{i u_2 y}+ C_3 e^{-i u_2 y} \qquad -b\leq y \leq 0 \\
C_4 e^{-i u_1 y}  \hspace{3cm} y\leq -b
\end{cases}\\
		\tilde{H}^{TE}_z=&\begin{cases}
\frac{C_5}{\eta_0} e^{iu_1y} \hspace{3cm} y\geq0\\
\frac{C_6}{\eta_0}e^{i u_2 y}+ \frac{C_7}{\eta_0} e^{-i u_2 y} \qquad -b\leq y \leq 0 \\
\frac{C_8}{\eta_0} e^{-i u_1 y}  \hspace{3cm} y\leq -b
\end{cases}.
	\end{align} 
	where $u_1$ is defined such that it has positive imaginary part and $C_1$ to $C_8$ are constants yet to be determined. We introduced the free space impedance $\eta_0=\sqrt{\mu_0/\epsilon_0}$ so all constants have the same dimensions. 
	
	At the interface between the dielectric substrate and air, the tangential components of $\vec{E}$ and $\vec{H}$ must be continuous. Introducing $A=-\omega\mu_0 I \frac{\gamma_1^2}{k_1^2}\frac{e^{i u_1 h}}{2u_1}$, these boundary conditions may be expressed in the following matrix equation
	\begin{equation}
		\underline{\underline{Q}}\begin{pmatrix}
		C_1 \\ C_2\\ C_3\\ C_4\\ C_5\\ C_6\\ C_7\\ C_8
		\end{pmatrix}=
		\begin{pmatrix}
		A\\0\\0\\0\\-\frac{k_1 \sqrt{\varepsilon_1}u_1}{\gamma_1^2}A\\0\\\frac{\beta \xi}{\gamma_1^2}A\\0
		\end{pmatrix}\label{eq:matrix}
	\end{equation}
	with relative dielectric permittivities $\varepsilon_j=\epsilon_j/\epsilon_0$ and matrix $\underline{\underline{Q}}$ which can be found in the appendix. $\underline{\underline{Q}}$ may be inverted to find expressions for the constants $C_1$ to $C_8$. We note here that the TE and TM modes are coupled through the continuity of $H_x$ and $E_x$ at the boundaries between substrate and air. Pure TM solutions with $C_5, C_6, C_7$ and $C_8$ equal to zero cannot fulfil the matrix equation. Hence, the resultant electromagnetic field will be hybrid in nature contrary to the Goubau mode for the dielectric coated cylinder. 
	
	\section{Characteristic Equation}
	So far we have discussed the exact solution for an arbitrary, infinitesimal current filament carrying a known current wave at a height $h$ above a substrate. However, in most practical cases, we want to find the propagation constant of the wave carried by an extended wire of finite size. This is still a formidable task even with the possibility of formally expressing the electromagnetic fields given any current distribution by convoluting the calculated fundamental solution with the source term \cite{Barton1989}.
	
	As a means of characterising the propagating mode, we introduce the effective refractive index $n_{eff}$ which is related to the propagation constant via $\beta=n_{eff}k_0$ . In order to formulate an approximate characteristic equation, we assume the wire is a perfect electrical conductor (PEC). This is valid for many metals in the GHz to THz frequency range if the radius is much larger than the skin depth. As a PEC, the tangential electric field should be zero at its surface. In particular, the $z$-component of the electric field must be zero. Imposing this condition at any point on the wire's surface, gives an equation for the approximate propagation constant if the wire is sufficiently thin \cite{Wait1972}. Hence, the characteristic equation may be expressed as
	\begin{equation}
	E_z(x=a, y=h)=0 \label{eq:Char}
	\end{equation}
	
	In a study on the validity of this approach for a wire in air above a semi-infinite earth, Pogorzelski and Chang showed that reasonable results are obtained if the contribution of azimuthal currents in the wire can be neglected \cite{Pogorzelski1977}. Furthermore, it was shown in the same work that given $k_2 h \ll 1$ and $|\gamma_1| h\ll1$ the contribution due to the first order azimuthal terms in the effective refractive index of the wave scales as 
	\begin{equation}
	\frac{\Delta n_{eff}}{n_{eff}^{(0)}}=\frac{g(n_{eff})}{n_{eff}^{(0)}\partial E_z/\partial n_{eff}|_{x^2+(y-h)^2=a^2}}\Big|_{n_{eff}=n_{eff}^{(0)}} \label{eq:higherOrder}
	\end{equation}
	where $\Delta n_{eff}$ is the correction due to higher order terms, $n_{eff}^{(0)}$ is the zero order effective index and $g(n_{eff})$ is a function of the effective refractive index and the geometry given by
	\begin{equation}
	g(n_{eff})=n_{eff}^2\Big(\frac{a}{2h}\Big)^2\Big(\frac{\epsilon_2-\epsilon_0}{\epsilon_2+\epsilon_0}\Big)\Big[\Big(\frac{a}{2h}\Big)^2+\frac{1}{1-2i\frac{ \epsilon_2-n_{eff}^2\epsilon_0}{(1-n_{eff}^2)(\epsilon_2+\epsilon_0)}}\Big]^{-1}
	\end{equation}
	Hence, the correction to the effective refractive index $\Delta n_{eff}$ due to azimuthal currents can be neglected if the absolute value of the right hand side in Equation \eqref{eq:higherOrder} is small. In our case the higher order contributions should be even smaller because the dielectric is only of finite thickness. Therefore, Equation \eqref{eq:higherOrder} provides an estimate on the obtainable accuracy when using the thin wire approximation to calculate the propagation constant. 
	
	In order to express and solve the characteristic equation, we focus on the amplitude $C_1$ which is required for calculating $E_z$ in real space above the substrate via the inverse Fourier transform \eqref{eq:Fourier}. Solving the matrix equation \eqref{eq:matrix}, we find that it may be written in the form
	\begin{equation}
		C_1=A(-1+F(\xi))
	\end{equation}
	where $F$ is a complicated function of $\xi$ whose complete form is given in the appendix. It has some noteworthy properties. First it only depends on $\xi^2$ reflecting the mirror symmetry of the system with respect to the plane $x=0$. Furthermore, for $|\xi|\gg |\gamma_2|$ and $|\xi b| \gg 1$ it asymptotically behaves  as
	\begin{equation}
	F(\xi)\sim \frac{2k_1^2}{\gamma_1^2}\frac{\xi^2 + u_1u_2}{k_1^2u_2 + k_2^2u_1}u_1 \sim \frac{k_1^2}{\gamma_1^2}\Big(1-\frac{2\beta^2}{k_1^2+k_2^2}\Big) \label{eq:F-asymptotic}
	\end{equation}
	which can be shown as all exponential terms $e^{iu_2b}$ in $F(\xi)$ will be very small. This expression is identical to the one obtained by Wait\footnote{Taking into account the different definitions of $u_1$ and $u_2$ used by Wait} for the case of a wire above a semi-infinite earth \cite{Wait1972}. In fact, in the corresponding limit of $b\rightarrow \infty$ the conditions on $\xi$ can be relaxed. This is due to all contributions involving $e^{iu_2b}$ becoming infinitely fast oscillating or zero so that they can be neglected in the inverse Fourier transform in eq. \eqref{eq:Fourier}. 
	
	 Let us reiterate the Fourier transform of $E_z$, which is of the following form 
	\begin{equation}
			\tilde{E_z}=-\omega\mu_0I\frac{\gamma_1^2}{k_1^2}\Big(\frac{e^{iu_1|y-h|}}{2u_1}-\frac{e^{iu_1(y+h)}}{2u_1}+\frac{e^{i u_1 (y+h)}}{2u_1}F(\xi)\Big)
	\end{equation}
	To calculate the inverse Fourier transform, we use the identity 
	\begin{equation}
		\int_{-\infty}^{\infty}\frac{e^{iu_1|(y\pm h)|}}{2u_1}e^{-i\xi x}d\xi=-iK_0\big(-i\gamma_1 \sqrt{x^2+(y\pm h)^2}\big)
	\end{equation}
	where $K_0$ is the zeroth order modified Bessel function of the second kind\cite{Gradshtein}. Furthermore, for a real number $V$ for which $V b \gg 1$ and $V\gg |\gamma_2|$ we may replace $u_1$ with $i\xi$ and $F(\xi)$ with its asymptotic form. Hence, the inverse transform gives 
	\begin{equation}
	\int_{V}^{\infty}F(\xi)\frac{e^{i u_1 (y+h)-i \xi x}}{2u_1} d\xi \sim\frac{-ik_1^2}{2\gamma_1^2}\Big(1-\frac{2\beta^2}{k_1^2+k_2^2}\Big)\Gamma\big(0,(ix+y+h)V\big)
	\end{equation}
	where $\Gamma$ is the incomplete Gamma function \cite{Abramowitz1972}. A similar expression is obtained for the integration from $-\infty$ to $-V$ as $F(\xi)$ only depends on $\xi^2$. In the region from -V to V, no analytic expression for the integral was found but it may be calculated numerically. Note that, in general, the Cauchy principal value of the integral needs to be taken because the integrand may contain poles.
	
	The field may then be expressed as
	\begin{multline}
	E_z=-i\omega \mu_0 I \frac{\gamma_1^2}{2\pi k_1^2}\Big[K_0\big(-i\gamma_1\sqrt{x^2+(y+h)^2}\big)-K_0\big(-i\gamma_1\sqrt{x^2+(y-h)^2}\big)\\-\frac{k_1^2}{\gamma_1^2}\big(1-\frac{2\beta^2}{k_1^2+k_2^2}\big)\Re\{\Gamma(0,(ix+y+h)V)\}-2i\int_{0}^{V}\cos(\xi x)F(\xi)\frac{e^{iu_1(y+h)}}{2u_1}d\xi\Big] \label{eq:Ez}
	\end{multline}
	where $\Re$ indicates that the real part of the expression in brackets should be taken. 
	
	\section{Solutions to the Characteristic Equation}
	
		\begin{figure}[t]
		\centering
		\includegraphics[width=0.48\textwidth]{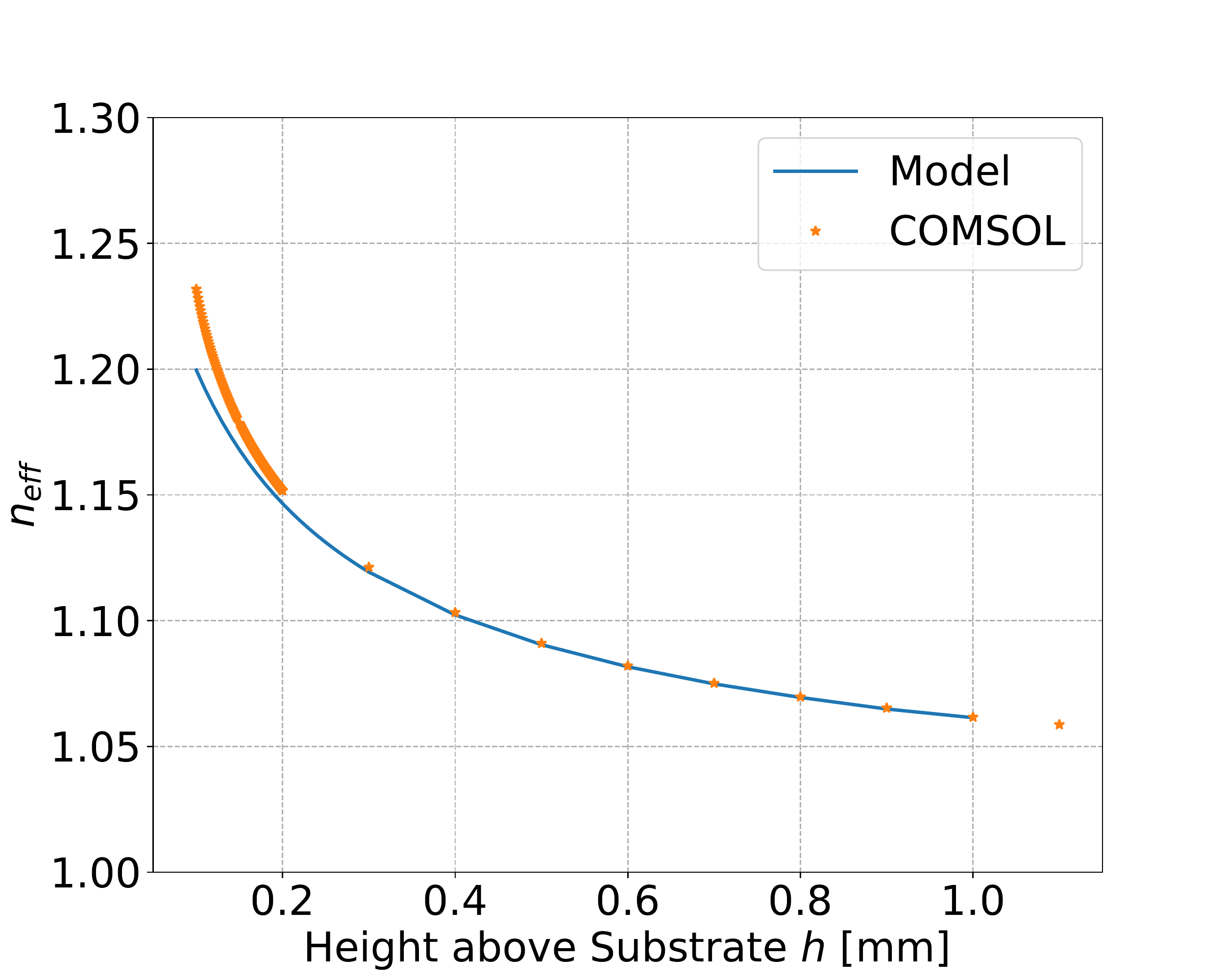}
		\includegraphics[width=0.48\textwidth]{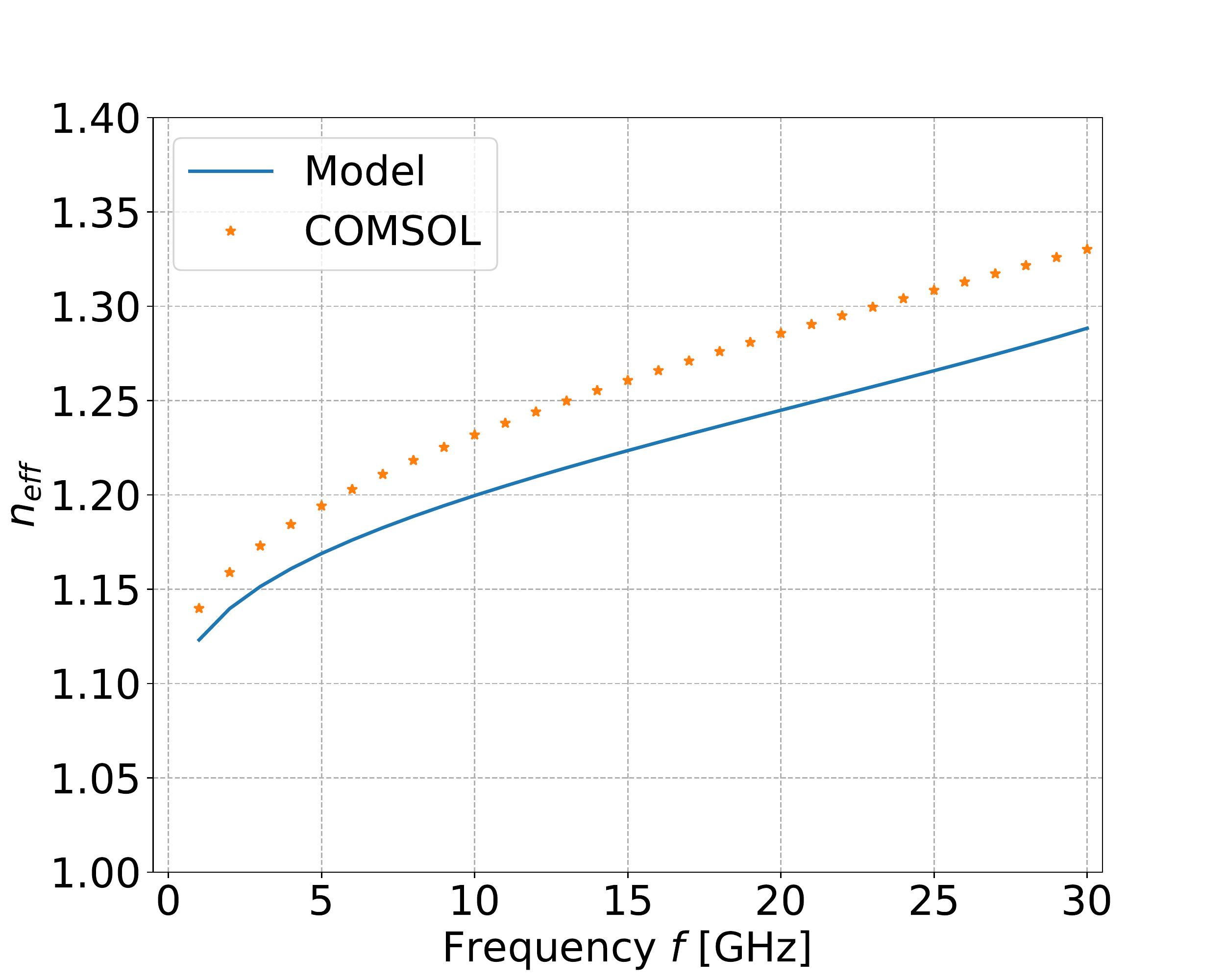}
		\includegraphics[width=0.48\textwidth]{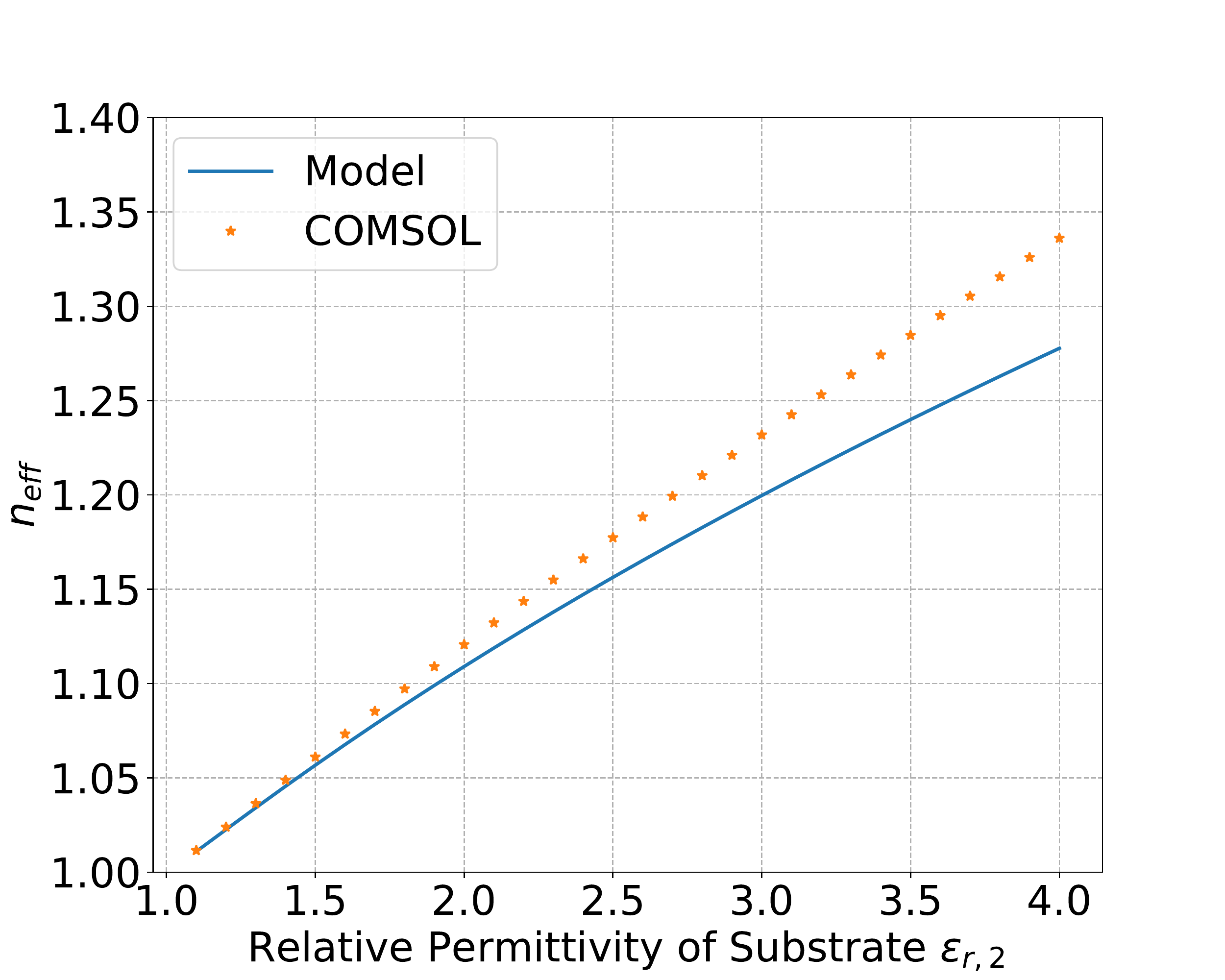}
		\includegraphics[width=0.48\textwidth]{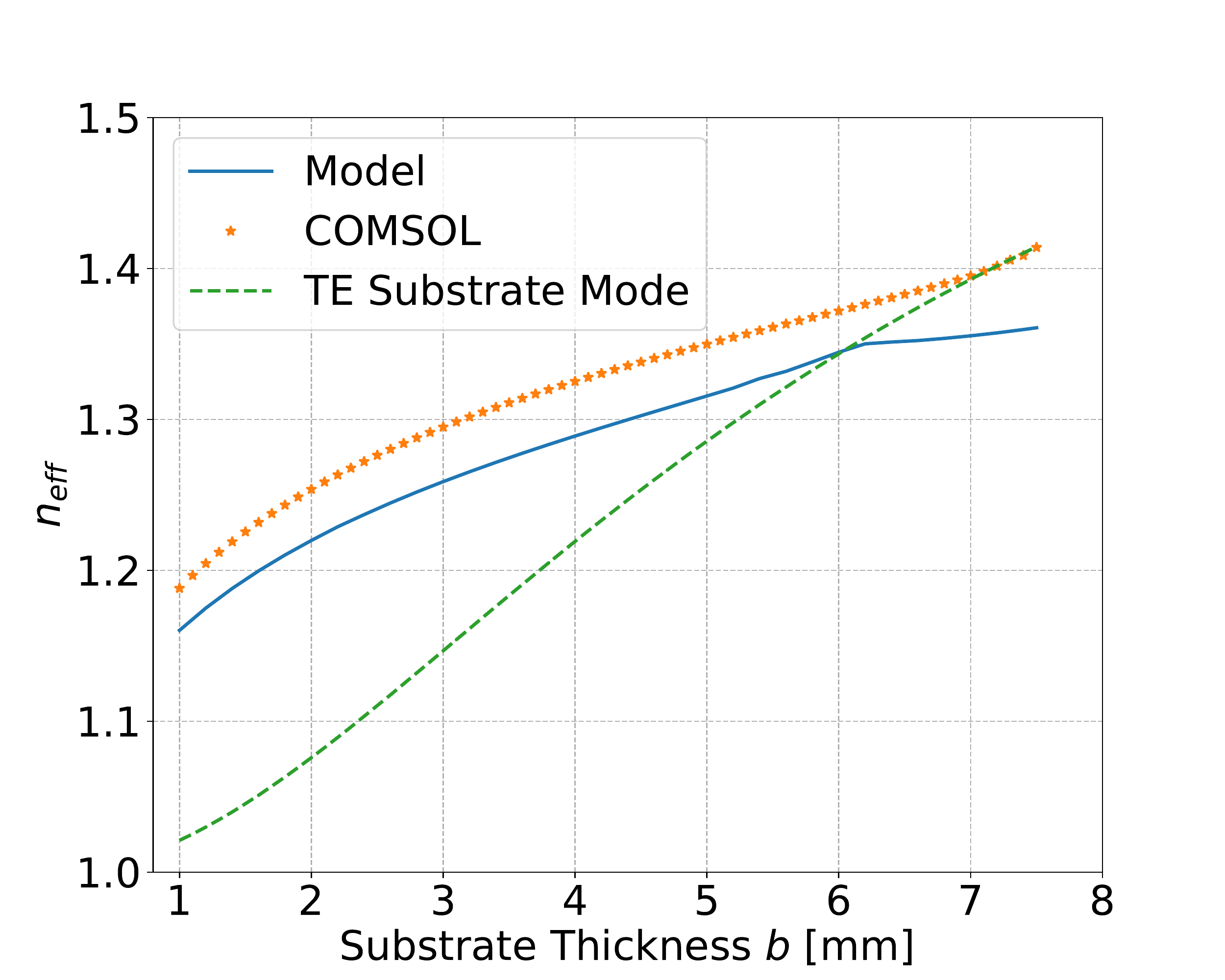}
		\caption{Sweep of different parameters showing their effect on the effective refractive index $n_{eff}$ of the propagating mode obtained with the presented model and the finite element solver Comsol Multiphysics. While sweeping one parameter, all other parameters were kept at their nominal values $a=0.1$~mm, $b=1.6$~mm, $\varepsilon_2=3$, $h=0.1$~mm and $f=10$~GHz.}
		\label{fig:sweep}	
	\end{figure}
	
	In general, the characteristic equation \eqref{eq:Char} must be solved numerically due to the integral containing $F(\xi)$. However, if the wire is located very far above the substrate such that $|\gamma_1 h| \gg 1$ , the integral may be neglected due to the strong exponential damping. In fact, the characteristic equation is then dominated by a single term
	\begin{equation}
		K_0(-i \gamma_1 a)=0
	\end{equation}
	This is the characteristic equation for a surface wave on a perfectly conducting cylinder surrounded by air which has been shown not to support any bound solutions \cite{Hondros1909}.
	
	 For all other cases we have solved equation \eqref{eq:Char} using Wolfram Mathematica. Due to the system being lossless, any bound mode will have a real $n_{eff}$ with $\sqrt{\varepsilon_1}<n_{eff}<\sqrt{\varepsilon_2}$ . Note that a large value of $n_{eff}$ generally indicates a stronger confinement of the wave to the wire and substrate. We examine the effects of the substrate thickness and dielectric constant, signal frequency and the wire's height above the substrate on the propagation constant by varying their values but keeping all other parameters constant. Nominal parameter values are  $a=0.1$~mm, $b=1.6$~mm, $\varepsilon_1=1$, $\varepsilon_2=3$, $h=0.1$~mm and frequency $f=10$~GHz (cf. Fig \ref{fig:setup}). Our results are validated against finite element numerical solutions obtained with COMSOL Multiphysics$^{\textregistered}$ \cite{COMSOL-AB}. Details on the simulation are given in the methods section. Figure \ref{fig:sweep} shows the results of the parameter sweeps. It can be seen from the figure that the effective refractive index and in turn the propagation constant crucially depends on all input parameters.
	
	For instance, the height of the wire above the substrate influences how much electromagnetic energy can travel inside the dielectric substrate. In general, if  the conductor is further away from the substrate, less energy travels in the dielectric. This means that the propagating mode has an effective refractive index closer to that of the surrounding medium. Consequently, as the conductor approaches the substrate, the effective refractive index increases as shown in Fig. \ref{fig:sweep}. Good agreement between our model and results obtained with Comsol can be seen. However, at small heights Comsol produces an effective index which is slightly higher than predicted by our method. 
	
	In fact, the results obtained with Comsol in Figure \ref{fig:sweep} seem to systematically lie above the results of our model. On the one hand, this deviation may be explained by our model neglecting the exact current distribution in the wire leading to errors such as those predicted in Equation \eqref{eq:higherOrder} which were on the order of 1-3\% throughout the sweep. On the other hand, as the height becomes very small, Comsol is forced to use elongated mesh elements between the wire and the substrate which is generally not recommended. 
	
	The effect of frequency on the PGL mode is shown in the second plot of Fig. \ref{fig:sweep}. At high frequencies the wave localises close to the conductor similar to the classical Sommerfeld or Goubau line. This leads to an increase in $n_{eff}$ which ensures a fast transverse decay in the surrounding air. Additionally, one can think of increasing the frequency as localising more of the wave energy inside the substrate which slows down the wave. Thus, the PGL is generally dispersive. At very low frequencies, the wire cannot be approximated as a PEC any more as the skin depth becomes similar to the wire radius. Hence, only values of the refractive index above 1 GHz are reported. Results obtained with our model and Comsol are within a few percent and both show the same general trend although Comsol predicts a slightly higher mode index.
	
	Increasing the dielectric constant of the substrate slows the wave down. This effect looks to be nearly linear in the magnitude of the effective dielectric constant. Agreement between the model and Comsol is good at small permittivities but the results deviate increasingly with the dielectric constant of the substrate. 
	
	 Finally, varying the substrate thickness influences the amount of energy that travels inside the substrate. Increasing the thickness slows the wave down leading to a higher effective refractive index. Figure \ref{fig:sweep} also includes the effective refractive index of the TE$_0$ substrate mode as Gacemi \textit{et al.} reported that with increasing substrate thickness the PGL mode mixes and ultimately merges into this mode \cite{Gacemi2013a}. Indeed the Comsol result aligns nicely with the TE$_0$ mode and past a thickness of 7.5~mm only the substrate mode was detected in Comsol. On the other hand, our model produces results beyond this thickness. However, a noticeable change in the behaviour $n_{eff}$ with the substrate thickness is observed after intersecting with the $n_{eff}$ of the substrate mode at around 6~mm. In fact, we were not able to obtain physical field patterns for values of the refractive index after this point. Hence, we believe that while results for larger substrate thicknesses can be calculated they do not hold any physical relevance. 
	
	Due to the dependence on geometrical parameters, the planar Goubau line can be designed to have a high or low effective refractive index signifying a strongly or weakly confined mode respectively. Clearly, a mode which is more localised near the substrate will experience increased dielectric loss. Hence, a trade-off between loss and field extent will need to be made. This can to some degree be mitigated by using low-loss substrates for instance quartz or plastics.

	\section{Field Pattern}
	
	\begin{figure}[t]
		\centering
		\raisebox{3mm}{\includegraphics[width=0.49\textwidth]{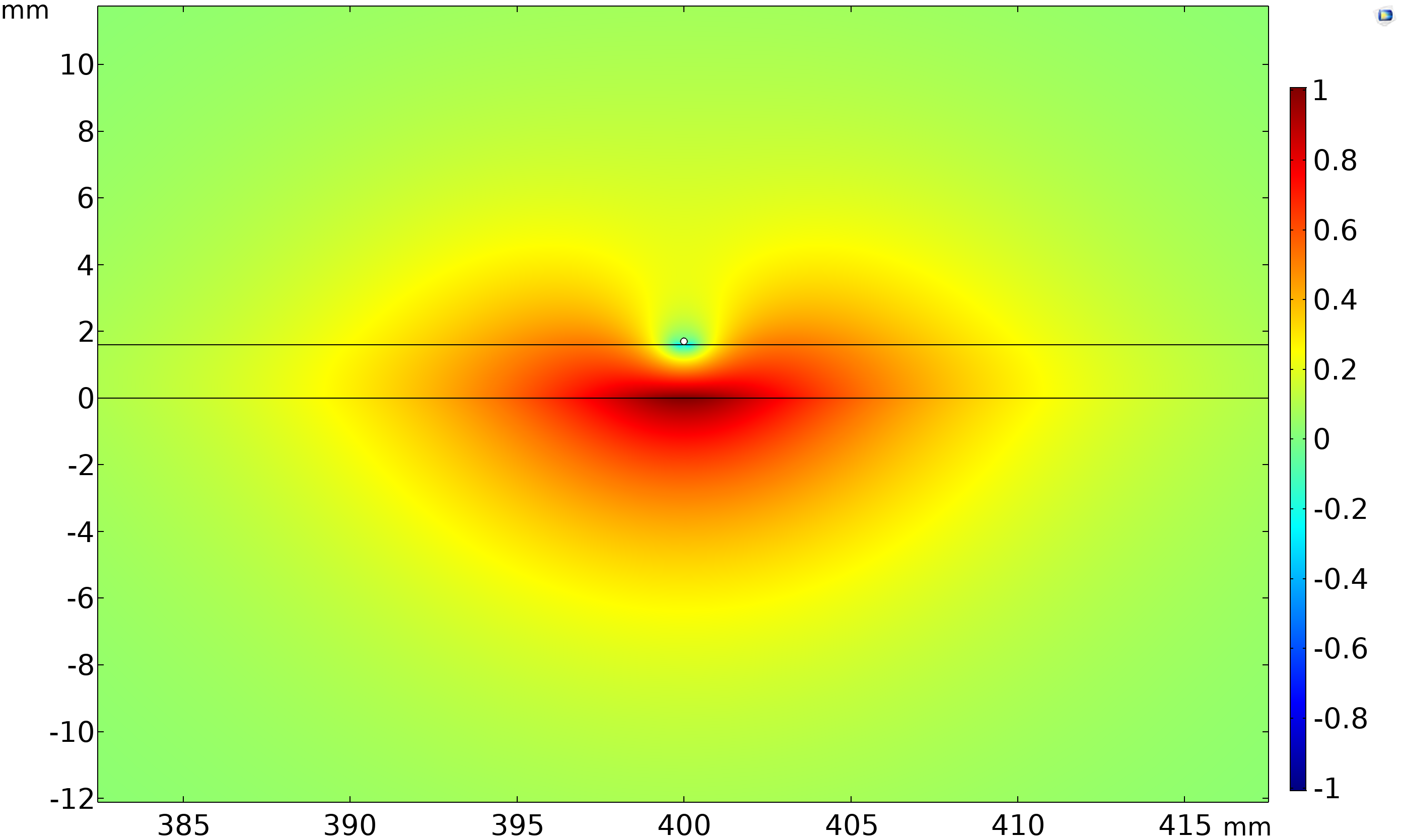}}
		\includegraphics[width=0.49\textwidth]{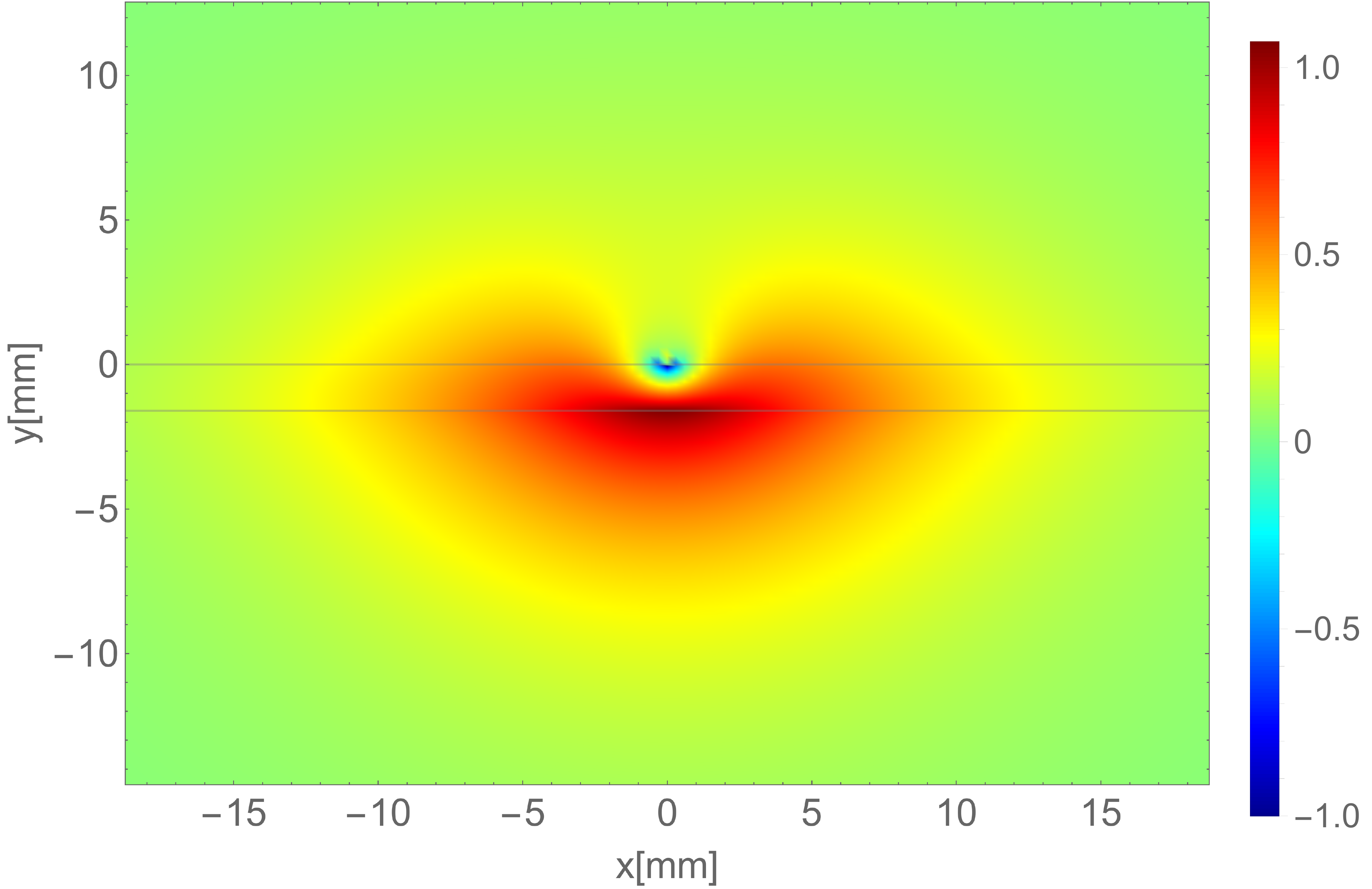}
		\includegraphics[width=0.45\textwidth]{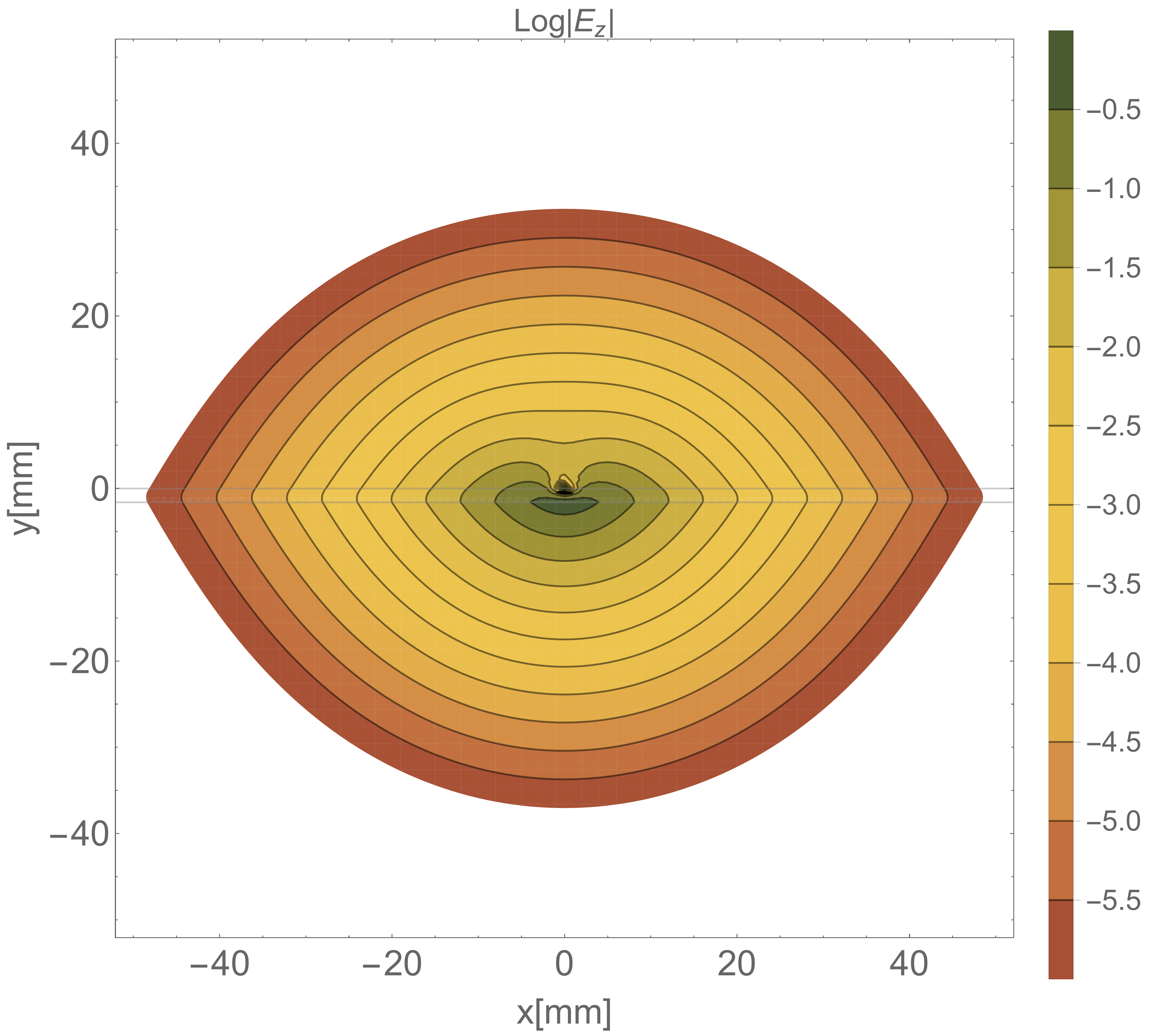}
		\caption{\textit{Top Left:} Plot of $E_z$ obtained through Comsol. \textit{Top Right:} Plot of $E_z$ calculated with the presented analytic model. Excellent agreement between the two field profiles is observed. \textit{Bottom:} Contour plot of $\log(|E_z|)$. The field is elongated along the substrate and exponentially decays away from the wire.}
		\label{fig:Ez}	
	\end{figure}
	
	\begin{figure}[t]
		\centering

	\end{figure}
	
	Once the characteristic equation has been solved, equation \eqref{eq:Fourier} may be computed at every point in space to calculate the distribution of the z-component of the electric field. Plots of the resulting field are shown in Fig \ref{fig:Ez} using the parameters given in the previous section. The top two plots show the longitudinal electric field calculated with Comsol and with our model. Both plots are in excellent agreement with each other. It is interesting to observe that the field changes sign between opposite sides of the dielectric. This behaviour was also found in the simulations presented by Horestani \textit{et al.} for the PGL but was not discussed \cite{Horestani2013}. We emphasise that the sign change is unique to the PGL and not found for the classic Goubau line where the mode is cylindrically symmetric. This shows that despite some similarities between the PGL and the classic Goubau line such as an exponential decay at large distances, the presence of a single sided substrate substantially alters the Goubau mode. It breaks the cylindrical symmetry of the system which in turn means no pure TM mode can propagate. As a result only a hybrid mode exists on the PGL. 
	
	The logarithmic contour plot at the bottom of Figure \ref{fig:Ez} shows that $E_z$ decays exponentially away from the wire at distances much greater than the wire radius. This can also be shown directly from our expression for $E_z$ in equation \eqref{eq:Ez}. There, we can neglect the integral for $|\gamma_1 y|\gg1$ due to the exponential damping. As the incomplete Gamma function is small for large argument, the field is dominated by the modified Bessel functions which have an exponential decay for large, real argument. Thus, the field drops off exponentially with a decay constant $|\gamma_1|$. This profile is consistent with the ones reported for planar Goubau lines in Refs. \cite{Horestani2013, Sanchez-Escuderos2013,Tang2017}.

	\section{Experimental Validation}

	 \begin{figure}[t]
	 	\centering
	 	\includegraphics[width=0.6 \textwidth]{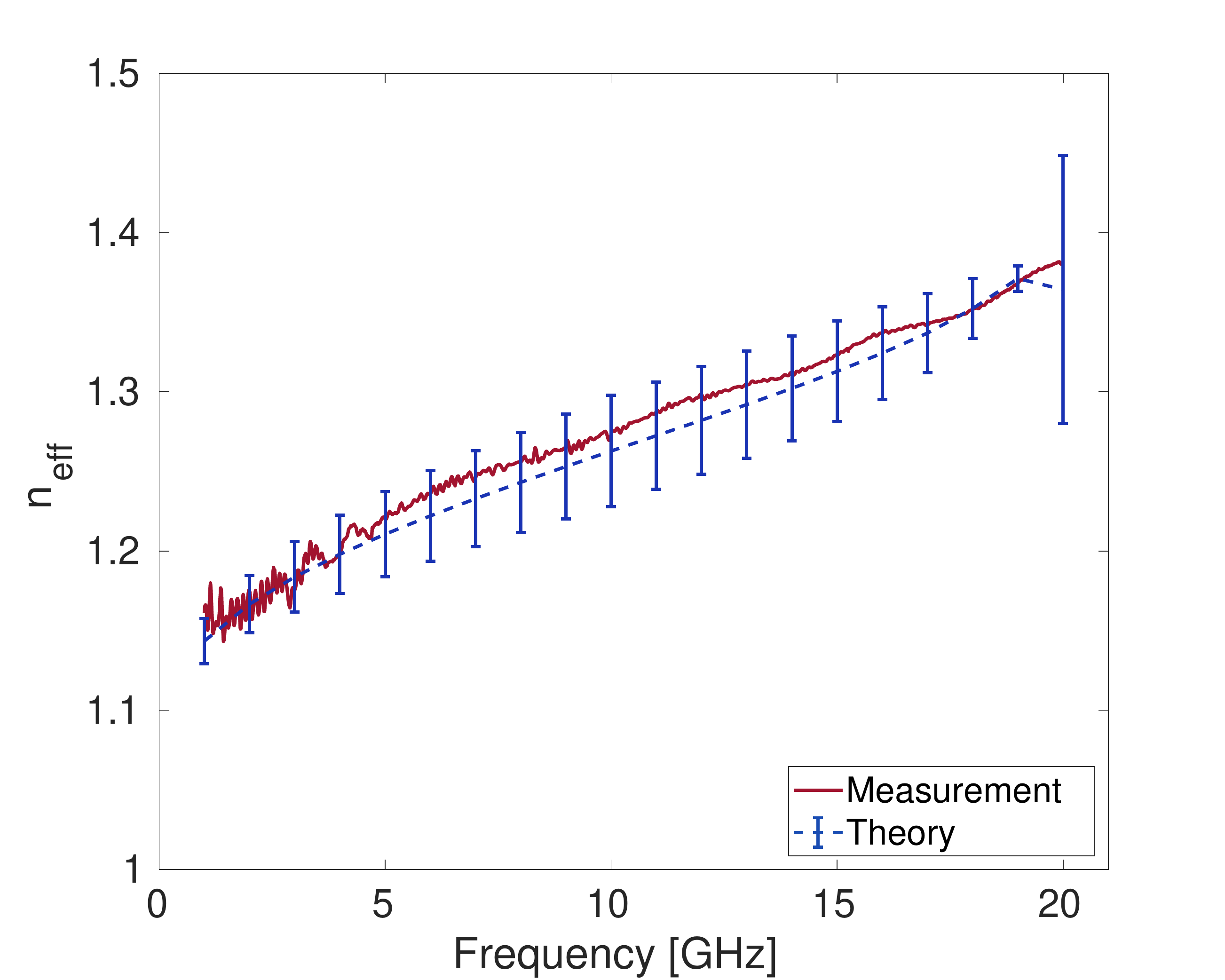}
 	\caption{Measured and theoretical effective refractive index of the PGL mode over a FR4 substrate. Error bars in the theory correspond to the uncertainty due to first order azimuthal currents as given in eq. \eqref{eq:higherOrder}.}
 	\label{fig:Measurement}
	 \end{figure}
	To validate our results experimentally, we measured the effective refractive index of the PGL mode as a function of frequency. This is achieved with a simple setup. Using scaled versions of the planar launchers discussed by Akalin \textit{et al.}, we excite a Sommerfeld surface wave on a single copper wire \cite{Akalin2006, Sommerfeld1899}. S-Parameters are obtained with a Vector Network Analyser (VNA). Then, we introduce a dielectric slab of finite length $l$ into the path and suspend the wire onto it using tape such that the dielectric represents the substrate discussed in our model. In transmission, the substrate will cause a phase delay $\Delta \varphi$ due to the increased refractive index of the now propagating PGL mode relative to the Sommerfeld mode. The delay can be measured with the VNA. Under the assumption that the Sommerfeld wave travels approximately at the speed of light, the phase delay is given by 
	\begin{equation}
		\Delta \varphi=k_0 l \big(n_{eff}-1\big)
	\end{equation}
	 This is now easily solved to give the effective refractive index of the PGL mode. Note that as we only measure the difference in phase that is introduced by the substrate, this measurement is independent of SW launching as long as a SW is propagating.
	 
	 The measurement results are shown in Figure \ref{fig:Measurement} together with theoretical predicted values. Error bars were added to the theoretical values according to equation \eqref{eq:higherOrder} estimating the higher order azimuthal current effects. Excellent agreement between theory and measurement is observed. Above 20~GHz the effect of azimuthal currents increases drastically reducing the accuracy of the presented theory. Error estimates for the measurement were omitted in the figure for clarity as they were significantly smaller than the theoretical ones.
	 
	\section{Considerations for Realistic Planar Goubau Lines}
	
	The presented simplified system has many similarities with the standard planar Goubau line such as the reported scaling behaviours with geometrical parameters and frequency. One difference is that in a real system, conductor and dielectric will be lossy leading to a complex effective refractive index. However, the main differences are that the conductor of a standard planar Goubau line has a rectangular cross section and the substrate is of finite width. In many cases the width of the substrate can be neglected as the electromagnetic fields of the bound Goubau mode decay exponentially. Thus, the edges do not have a strong influence on the field pattern. However, the shape of the conductor has been shown to strongly influence the effective refractive index \cite{Gacemi2013}. 
	
	It is beyond the scope of this work to derive a complete theory to incorporate these effects. However, we will try to give some qualitative arguments to describe the observed trends within our framework. If we consider a conductor of finite thickness but variable width, then most of the electric fields and currents will be localised near the edges due to the lightning rod effect. Hence, a natural model is two parallel wires on the substrate which are located at the edges of the rectangular conductor carrying coupled surface waves. This type of coupling has recently been studied for Sommerfeld wires and it was shown to reduce the effective refractive index \cite{Molnar2020}. This behaviour is consistent with the observations in Ref. \cite{Gacemi2013}.

	\section{Conclusion}

	This paper presents a theoretical investigation of a simplified Goubau line consisting of a cylindrical wire above an infinitely wide substrate. To this end, the electromagnetic field of an infinitesimal current filament above a substrate has been derived in Fourier space. By incorporating the finite width of a realistic wire, an approximate characteristic equation was derived. Estimates to the applicability of this equation were given before exploring the influence of  geometrical parameters and frequency on the wave characteristics. It was found that, depending on the operating frequency and setup, both weakly and strongly confined surface waves can propagate allowing one to tune the geometry depending on the desired application. Furthermore, the field profile for a particular set of parameters was calculated. All results agreed well with numerical simulations and previously reported experimental studies. Finally, the derived model was experimentally tested and excellent agreement between theory and measurement was found. 
	
	This work is one of the first attempts at analytically modelling the planar Goubau line. Although only a simplified version was discussed, the behaviour with frequency and other parameters was found to agree with the previous reports for a standard planar Goubau line with rectangular conductor. Hence, it may be considered as an important step towards a better understanding of the PGL. The limitations of the presented model with respect to a realistic planar Goubau line with rectangular conductor were discussed in the last section and some qualitative arguments were made to incorporate the effect of the conductor geometry. In conclusion, we expect these insights to help better understand and utilise planar Goubau lines in printed circuit board designs for challenging applications such as terahertz spectroscopy or high frequency circuitry. 
	
	\paragraph*{Methods} 	In our experimental setup a 15~cm long, 1.6~mm thick, dielectric slab made from FR4 epoxy in conjunction with a 0.5~mm annealed copper wire were used. The relative permittivity of the substrate is given as 4.55 by the manufacturer. S-parameters were measured by a 8722D VNA from Agilent Technologies.
	
	\paragraph{Author Contributions}
	TS conceived the theoretical model and any calculation based on it. Furthermore, he carried out the experiment. DM was responsible for numerical modelling and simulation with Comsol. Additionally, DM and TS jointly interpreted the resulting data and drafted the manuscript. AM and MP crucially revised the manuscript and supervised the research. All authors approve the current version of the manuscript.
	
	\paragraph{Competing Interests}
	The authors declare no competing interests.
	
	 \paragraph*{Acknowledgements}
	 
 	This work was supported by the Royal Society Grants IF170002 and INF-PHD-180021. Additional funds were provided by BT plc and Huawei Technologies Co., Ltd. The authors thank the Royal Society, BT and Huawei for these funds.

	\bibliographystyle{naturemag}

	\appendix
	\section{Matrix \underline{\underline{$Q$}}}
	\begin{multline*}
	\underline{\underline{Q}}=	\left(
	\begin{matrix}
		-1 &1&1&0\\
		0&e^{-iu_2b}& e^{iu_2b}& -e^{iu_1b}\\
		0&0&0&0\\
		0&0&0&0\\
		-\frac{k_1 \sqrt{\varepsilon_1}u_1}{\gamma_1^2}&\frac{k_2 \sqrt{\varepsilon_2}u_2}{\gamma_2^2}&-\frac{k_2 \sqrt{\varepsilon_2}u_2}{\gamma_2^2}&0\\
		0 & \frac{k_2\sqrt{\varepsilon_2} u_2}{\gamma_2^2}e^{-iu_2b}&-\frac{k_2\sqrt{\varepsilon_2} u_2}{\gamma_2^2}e^{iu_2b}&\frac{k_1 \sqrt{\varepsilon_1} u_1}{\gamma_1^2}e^{iu_1b}\\
		-\frac{\beta \xi}{\gamma_1^2}&\frac{\beta \xi}{\gamma_2^2}&\frac{\beta \xi}{\gamma_2^2}&0\\
		0&\frac{\beta \xi}{\gamma_2^2}e^{-iu_2b}&\frac{\beta \xi}{\gamma_2^2}e^{iu_2b}&-\frac{\beta \xi}{\gamma_1^2}e^{-iu_1b}
	\end{matrix} \right. \\
	\left. \begin{matrix}
	0&0&0&0\\
	0&0&0&0\\
	1&-1&-1&0\\
	0&e^{-iu_2b}&e^{iu_2b}&-e^{iu_1b}\\
	-\frac{\beta \xi}{\gamma_1^2}&\frac{\beta \xi}{\gamma_2^2}&\frac{\beta \xi}{\gamma_2^2}&0\\
	0&\frac{\beta \xi}{\gamma_2^2}e^{-iu_2b}&\frac{\beta \xi}{\gamma_2^2}e^{iu_2b}&-\frac{\beta \xi}{\gamma_1^2}e^{iu_1b}\\
	\frac{k_0 u_1}{\gamma_1^2}&-\frac{k_0 u_2}{\gamma_2^2}&\frac{k_0 u_2}{\gamma_2^2}\\
	0&-\frac{k_0 u_2}{\gamma_2^2}e^{-iu_2b}&-\frac{k_0 u_2}{\gamma_2^2}e^{iu_2b}&-\frac{k_0 u_1}{\gamma_1^2}e^{iu_1b}
	\end{matrix} \right)
	\end{multline*}
	\section{Function $F(\xi)$}
	\begin{gather*}
		F(\xi)=2 \gamma_2^2 k_0^2 u_1 \Big[-\beta^2 (-1 + e^{2 i u_2 b}) (\gamma_1^2 - 
		\gamma_2^2)^2 \xi^2 (\gamma_2^2 u_1 + \gamma_1^2 u_2) - \\
		k_0 (\gamma_2^2 k_0 u_1 + 
		\sqrt{\varepsilon_2} \gamma_1^2 k_2 u_2) \Big((-1 + e^{2 i b u_2}) \gamma_2^4 u_1^2 - 
		2 (1 + e^{2 i b u_2}) \gamma_1^2 \gamma_2^2 u_1 u_2 + \\
		(-1 + e^{2 i b u_2}) \gamma_1^4 u_2^2 \Big) + 
		e^{2 i b u_2} \Big(-\beta^2 (-1 + e^{2 i b u_2}) (\gamma_1^2 - 
		\gamma_2^2)^2 \xi^2 (-\gamma_2^2 u_1 + \gamma_1^2 u_2) +\\ 
		k_0 (\gamma_2^2 k_0 u_1 - \sqrt{\varepsilon_2} \gamma_1^2 k_2 u_2) ((-1 + e^{2 i b u_2}) \gamma_2^4 u_1^2 - 
		2 (1 + e^{2 i b u_2}) \gamma_1^2 \gamma_2^2 u_1 u_2 \\+ (-1 + e^{2 i b u_2}
		) \gamma_1^4 u_2^2)\Big)\Big]/\Big[\beta^4 (-1 + e^{2 i b u_2})^2 (\gamma_1^2 - \gamma_2^2)^4 \xi^4 +\\ 
		2 \beta^2 (-1 + e^{2 i b u_2}) (\gamma_1^2 - 
		\gamma_2^2)^2 k_0 \xi^2 ((-1 + e^{2 i b u_2}) \gamma_2^4 k_0 u_1^2 - \\(1 + e^{2 i b u_2}) \gamma_1^2 \gamma_2^2 (k_0 + \sqrt{\varepsilon_2} k_2) u_1 u_2 + (-1 + e^{2 i b u_2}) \sqrt{\varepsilon_2} g_1^4 k_2 u_2^2) + \\
		k_0^2 ((-1 + e^{2 i b u_2}) \gamma_2^4 u_1^2 - 
		2 (1 + e^{2 i b u_2}) \gamma_1^2 \gamma_2^2 u_1 u_2 + (-1 + e^{2 i b u_2}) \gamma_1^4 u_2^2)\\
		 \Big((-1 + e^{2 i b u_2}) \gamma_2^4 k_0^2 u_1^2 - 
		2 (1 + e^{2 i b u_2}) \sqrt{\varepsilon_2}
		\gamma_1^2 \gamma_2^2 k_0 k_2 u_1 u_2 + (-1 + e^{2 i b u_2}) \varepsilon_2 \gamma_1^4 k_2^2 u_2^2\Big)\Big]
	\end{gather*}
	
\end{document}